\documentclass[a4paper,11pt]{article}
\usepackage{pos,hyperref}
\newcommand{\amuhlo}{$a_{\mu}^\text{HLO}$}

\newcommand{\gmtwo}{\ensuremath{g\!-\!2}}
\newcommand{\amu}{$a_{\mu}$}

\title{Towards a full NNLO Monte Carlo generator for low energy $e^+e^-$ data into leptons and hadrons}

\author[a,b]{Graziano Venanzoni}

\affiliation[a]{University of Liverpool, Liverpool L69 3BX, United Kingdom}

\affiliation[b]{INFN Sezione di Pisa, Largo Bruno Pontecorvo 3, 56127, Pisa, Italy}

\emailAdd{graziano.venanzoni@liverpool.ac.uk}

\abstract{
During the last 15 years the Radio MonteCarLow 
Working Group has been providing valuable support to
the development of radiative corrections and Monte Carlo event generators for low energy $e^+e^-$ data and $\tau$-lepton decays. 
While the working group has been operating for more than 15 years without a formal basis for funding, parts of our program have recently been included as a Joint Research Initiative in the group application of the European hadron physics community, STRONG2020, to the European Union, with a more specific goal of creating an annotated database for low-energy hadronic cross sections in $e^+e^-$ collisions. 
In parallel the theory community is continuing its effort towards the realization of improved Monte Carlo generators with for low energy $e^+e^-$ data into hadrons. Full NNLO corrections in the leptonic sector are to be combined with an improved treatment of radiative corrections involving pions. This is of relevance for the precise determination of the leading hadronic contribution to the muon \gmtwo. We will report on these initiatives.}

\FullConference{The European Physical Society Conference on High Energy Physics (EPS-HEP2023)\\
 21-25 August 2023\\
Hamburg, Germany\\}

\begin{document}
\maketitle

\section{Introduction}
\label{sec:intro}
The importance of continuous and close collaboration between the experimental and theoretical groups is crucial in the quest for precision in hadronic physics. This is the reason why the Working Group (WG) on “Radiative Corrections and Monte Carlo Generators for Low Energies” (Radio MonteCarLow, see~\cite{radiomc}) was formed a few years ago bringing together experts (theorists and experimentalists) working in the field of low-energy $e^+e^-$ physics and partly also the $\tau$-lepton community. Its main motivation was to understand the status and the precision of the Monte Carlo generators (MC) used to analyze the hadronic cross section measurements obtained in energy scan experiments as well as  with radiative return method, to determine luminosities. Whenever possible specially prepared comparisons, {\it i.e.} comparisons of MC generators with a common set of input parameters and experimental cuts, were performed within the project. The main conclusions of this major effort were summarized in a report published in 2010~\cite{Actis:2010gg12}. During the years the WG structure has been enriched of more research topics including  seven subgroups: Luminosity, R-measurement, ISR, Hadronic Vacuum Polarization, muon \gmtwo\ and $\Delta\alpha(M_{Z^0})$, 2-photon physics, FSR models, $\tau$ decays.
The working group had been operating for more than 15 years without a formal basis and a dedicated funding. Recently parts of the program have been  included as a Joint Research Initiative (JRA3-PrecisionSM) in the group application of the European hadron physics community, STRONG2020 (\url{http://www.strong-2020.eu}), to the European Union, with a more specific goal of creating an annotated database for the low-energy hadronic cross section data in $e^+e^- $ collisions.
All these efforts have been recently revitalized by the new high-precision measurements of the anomalous magnetic moment of the muon by the Muon g-2 Collaboration in Fermilab~\cite{Muong-2:2021ojo,Muong-2:2023cdq} which show a more than $5\sigma$ discrepancy with respect to the state-of-the-art theoretical prediction from the Standard Model (SM) computed by the  Muon $g\!-\!2$ Theory Initiative  in 2020~\cite{Aoyama:2020ynmqq}. However a recent high-precision lattice evaluation of the BMW collaboration~\cite{Borsanyi:2020mff} shows tension with the time-like data-driven determinations of \amuhlo, being $2.1\sigma$ higher than the prediction from the dispersive approach used in~\cite{Aoyama:2020ynmqq}. In addition the CMD-3 Collaboration in Novosibirsk released in 2023 a new measurement of the $e^+e^-\to\pi^+\pi^-$ cross section~\cite{CMD-3:2023alj} that significantly disagrees with all previous measurements used in~\cite{Aoyama:2020ynmqq}, and, taken in isolation, gives a prediction of \amu\ in closer agreement with the experimental value~\cite{CMD-3:2023alj}.
In view of the  current tensions and puzzles in the hadronic sector and the experimental efforts under way at Fermilab (USA) and J-PARC (Japan)~\cite{Abe:2019thb} to improve the $a_{\mu}$ accuracy, a consolidation (and a possible improvement) of the SM prediction of the muon \gmtwo\ is extremely important. Here we discuss the status of the radiative corrections and Monte Carlo tools for low energy $e^+e^-$ data and the prospects towards the full NNLO MC calculation for the leptonic part, combined with improved treatment of radiative corrections with pions in the final state.

\section{Hadronic Vacuum Polarization to the Muon \gmtwo}
The theory prediction for the muon magnetic anomaly is limited by vacuum fluctuations involving strongly interacting particles. They mainly originate from the leading hadronic vacuum polarization term \amuhlo\ which cannot be reliably calculated perturbatively in QCD, due to the non-perturbative nature of the strong interactions at low energy. It is possible to overcome this problem by means of a dispersion relation technique involving experimental data measuring the cross section of electron positron annihilation into hadrons, $e^+e^-\to hadrons$, (so-called "time-like” or “dispersive” approach):

\begin{eqnarray}
      a_{\mu}^{\mbox{$\scriptscriptstyle{\rm HLO}$}} &=& 
      \frac{1}{4\pi^3}
      \int^{\infty}_{m_{\pi}^2} {\rm d}s \, K(s) \sigma^{0}\!(s) \nonumber \\
                                                                  &=&
      \frac{\alpha^2}{3\pi^2}
      \int^{\infty}_{m_{\pi}^2} {\rm d}s \, K(s) R_{had}(s)/s \, .
\label{eq:amu_had}
\end{eqnarray}
\noindent where $R_{had}(s)$ is the ratio of the total $e^+e^- \to hadrons$ and the Born $e^+e^- \to \mu^+\mu^-$ cross sections in the pointlike ($m_\mu=0$) limit,  $K(s)$ is a smooth function and $m_\pi$ is the pion mass. The functional form of the integral emphasizes low energy contributions where the cross section $e^+e^- \to hadrons$ is densely populated with resonances and modulated by threshold effects. This makes the dispersive approach evaluation of \amuhlo\ highly challenging with an error dominated by systematic uncertainties of data~\cite{Aoyama:2020ynmqq}.
\noindent Usually the computation of \amuhlo\ is done by using experimental data in the low-energy range and perturbative QCD  at higher energy.
The energy cut between the two regimes varies in different calculations. 
In the region of relatively high energies, inclusive measurements of the cross
section are carried out while at lower energies $<2$ GeV, the exclusive
measurement of cross sections of each separate final hadron state $e^+e^- \to 2 \pi; 3\pi; 4\pi; 2K;...$ is mainly used, and \amuhlo\
is calculated as a sum of contributions due to individual final hadron states.
The last 20 years have seen a big effort on $e^+e^-$ data in the low energy region. Better data were produced at fixed energy (CMD-3 and SND at VEPP-2000 and BESIII at BEPC colliders) and flavor factories where the use of Initial State Radiation (ISR), pioneered by the KLOE and BaBar experiments, opened a new way to precisely obtain the $e^+e^-$ annihilation cross sections into hadrons at particle factories operating at fixed beam energy. New dedicated tools and refined theoretical treatment were developed for the analysis of data~\cite{Actis:2010gg12}. All this effort led to a substantial reduction on the uncertainty of \amuhlo\ to 0.6\%~\cite{Aoyama:2020ynmqq}.

\section{Radiative corrections}
\label{sec:another}

The precise determination of the hadronic cross sections (accuracy $\lesssim 1\%$) 
requires an excellent control of higher order effects like radiative
 corrections and the 
non-perturbative hadronic contribution to the running of $\alpha$ 
(i.e. the vacuum polarisation, VP)
in MC programs used for the analysis of the data.
Particularly in the last years, the increasing precision reached 
on the experimental side at the $e^+e^-$ colliders (VEPP-2M, DA$\mathrm{\Phi}$NE, BEPC, PEP-II and KEKB)
led to the development of dedicated high precision theoretical tools:
BabaYaga (and its successor BabaYaga@NLO) 
for the measurement of the luminosity, 
MCGPJ for the simulation of exclusive leptonic and hadronic channels, and 
PHOKHARA for the simulation of the process  with Initial State Radiation (ISR) $e^+e^-\to hadrons+\gamma$,
are examples of MC generators which include NLO corrections with per mill accuracy.
In parallel to these efforts, well-tested codes such as 
BHWIDE (developed for LEP/SLC colliders) were adopted.
Theoretical accuracies of these generators were estimated, whenever possible, 
 by evaluating  missing higher order contributions. From this point of view, the great progress in the calculation of two-loop corrections to the Bhabha scattering cross section was essential to 
establish the high theoretical accuracy of the existing generators for the luminosity measurement.
However, usually only analytical or semi-analytical estimates
 of missing terms exist which don't take into account 
realistic experimental cuts.
In addition, MC event generators include different parameterisations for the VP
 which affect the prediction (and the precision) 
of the cross sections and also the radiative corrections are usually implemented differently. 
These arguments evidently imply the importance of comparisons of MC generators with a common set of input parameters and experimental cuts,
which allow to check that the details entering the complex structure of the generators are under control and free of possible bugs. 
Such {\it tuned} comparisons, which started in the LEP era, are a key step for the validation of the generators and were regularly discussed at the Radio MonteCarLow WG meetings~\cite{radiomc}.
\begin{table}[]
\begin{center}
\scalebox{.85}{
\begin{tabular}{ |c|c|c|c| } 
 \hline
 \multicolumn{4}{|c|}{MC generators for exclusive channels} \\
 \multicolumn{4}{|c|}{(exact NLO + Higher Order terms in some approximation)} \\
 \hline
 MC generator & Channel & Precision & Comment (see caption)\\ 
 \hline
MCGPJ (VEPP-2M, VEPP-2000) & $e^+e^-\to e^+e^-; \mu^+\mu^-; \pi^+\pi^-;...$ & 0.2\% & (A) \\
\hline
BabaYaga@NLO (KLOE, BaBar, BESIII)  & $e^+e^-\to e^+e^-; \mu^+\mu^-; \gamma \gamma$ & 0.1\% & (B) \\
 \hline
BHWIDE (LEP) & $e^+e^-\to e^+e^-$ & 0.1\% & (C) \\
\hline
\end{tabular}
}
\caption{MC generators for exclusive channels. (A) = photons jets using collinear structure functions with exact NLO matrix elements; (B) = QED Parton Shower approach with exact NLO matrix elements; (C) =  Yennie-Frautschi-Suura (YFS) exponentiation method with exact NLO matrix elements. Description of these MC generators can be found in~\cite{Actis:2010gg12}.}
\label{Tab1}
\end{center}
\end{table}

\begin{table}
\begin{center}
\scalebox{0.85}{
\begin{tabular}{ |c|c|c|c| } 
\hline
\multicolumn{4}{|c|} {MC generators for ISR} \\
\multicolumn{4}{|c|}{(from approximate to exact NLO)} \\
 \hline
 MC generator & Channel & Precision & Comment (see caption)\\ 
 \hline
EVA (KLOE) & $e^+e^-\to\pi^+\pi^-\gamma$ & O(\%) & (D) \\
\hline
AFKQED (BaBar) & $e^+e^-\to \pi^+\pi^-\gamma$ & Depends on event selection & (E) \\
 \hline
PHOKHARA & $e^+e^-\to \pi^+\pi^-\gamma;\mu^+\mu^-\gamma$ & 0.5\% & (F) \\
(KLOE, BaBar, BESIII) & $e^+e^-\to hadrons\ \gamma$ & O(\%) &  \\
\hline
KKMC & $e^+e^-\to l^+ l^- \gamma$ & High accuracy only for muon pairs & (G) \\
\hline
\end{tabular}
}
\caption{MC generators for ISR. (D) = Tagged photon; ISR at LO + collinear structure functions; FSR: sQED+Form Factor; (E) = ISR at LO + collinear structure functions; (F) = ISR and FSR (sQED+Form Factor) at NLO; (G) = Yennie-Frautschi-Suura (YFS) exponentiation  for soft photons + hard part and subleading terms in some approximation. Description of these MC generators can be found in~\cite{Actis:2010gg12}.}
\label{Tab2}
\end{center}
\end{table}

\section{Prospects on radiative corrections and MC tools}
With a lot of new data from VEPP-2000, BaBar, Belle II, and BESIII experiments with better quality and refined systematic errors,  radiative correction and MC generators (including ISR) should aim at 0.1\% uncertainty. 
\noindent Tables~\ref{Tab1} and~\ref{Tab2}  show the status of MC generators for exclusive channels and for ISR respectively.
All these generators have matrix elements which go from approximate to exact NLO. To reach 0.1\% uncertainty a  MC generator for low-energy $e^+e^-$  cross section into hadrons with NNLO corrections for leptonic part, combined with improved treatment of radiative corrections with pions in the final state would be required. In the last years a renovated effort~\cite{Abbiendi:2022liz} has started on radiative corrections and Monte Carlo tools for low-energy hadronic cross sections in $e^+e^-$ collisions with comparison between different generators and refined models for the treatment of hadron-photon interaction. New analyses at BaBar~\cite{BaBar:2023xiy} and CMD-3~\cite{Ignatov:2022iou} have shown the importance of improved treatment of radiative corrections and hadron-photon modelization. Impact of missing NLO radiative corrections in the KLOE analysis has been studied~\cite{Campanario:2019mjh, mueller}.
A new dedicated effort for the improvement of the radiative corrections and Monte Carlo tools for 
    low-energy hadronic cross sections in $e^+e^-$ 
    collisions has started within the STRONG2020 activity. A  WorkStop/ThinkStart has been organized at the
University of Zurich in June 2023 where the activity was divided into 5 packages~\cite{workstop}: (1) leptonic processes at NNLO; (2) $\gamma^* \to$ leptons at NNNLO; (3) processes with hadrons; (4) parton showers; (5)  experimental guidance.

\noindent In view of this and the parallel effort on the NNLO MC generator for $\mu-e$ elastic scattering process for the MUonE experiment~\cite{Banerjee:2020tdt}, the goal of a MC generator for low-energy $e^+ e^-$  cross section into hadrons with NNLO corrections for leptonic part, combined with improved treatment of radiative corrections with pions in the final state, although very challenging may be feasible in the next few years.
\section{The PrecisionSM database}
\label{strong2020}
 The PrecisionSM database~\cite{precisionsm} is one of the specific objectives of STRONG2020, a European joint research initiative that groups together researchers from different scientific frontiers (low energy, high energy, instrumentation and infrastructures) with the broad goal of study strong interactions and  develop applications for beyond fundamental physics. It contains 
 information about the reliability of the data sets, their systematic errors and the treatment of radiative corrections. 
The status of the database is discussed in~\cite{adriutti}.

\section{Conclusion}
The importance of continuous and close collaboration between the experimental and theoretical groups is crucial in the quest for
precision in hadronic physics.  That was done in the last 15 years within the  Working Group on "Radiative Corrections
and Monte Carlo Generators for Low Energies" (Radio MonteCarLow). That effort resulted in an accuracy on MC generators for low energy $e^+e^-$ processes between 0.2 and 0.5\%. New data and improved evaluation of \amuhlo\ requires
improvement on MC generators at ~0.1\% which would require a NNLO matrix elements calculation for the leptonic part combined with improved treatment of radiative corrections with pions in the final state. That accuracy although challenging can be feasible within few years~\cite{Abbiendi:2022liz,workstop}. In addition to the request in accuracy, for the better understanding of the experimental data and to facilitate the information, the STRONG2020 project will contribute with an annotated database for low energy
hadronic cross sections in $e^+e^-$ collisions, which will contain information about the reliability of the data sets, their systematic errors, and the treatment of radiative corrections.

\section*{Acknowledgements}
This work was supported by the European Union STRONG2020 project under Grant Agreement Number 824093 and by the Leverhulme Trust, LIP-2021-01. I would like to thank my colleagues A. Driutti, F. Ignatov, A. Kupsc, A. Lusiani, S. M\"uller, A. Signer, T. Teubner and Y. Ulrich for a common activity within STRONG2020, and H. Czy\.z for useful comments.



\end{document}